\documentclass[sigconf,twocolumn, natbib=false, anonymous=false]{acmart}

\setcopyright{none}
\acmISBN{}
\acmConference{Submitted for review}{July 26}

\usepackage{amsmath,amsfonts}
\usepackage{algorithmic}
\usepackage{graphicx}
\usepackage{hyperref}
\usepackage{textcomp}
\usepackage[backend=biber, style=numeric, datamodel=acmdatamodel, sorting=none]{biblatex}
\usepackage{multirow}
\usepackage{tabularx} 
\usepackage{caption}
\usepackage{booktabs}
\usepackage{wrapfig}
\usepackage{enumitem}
\usepackage{listings}
\usepackage{acronym}
\usepackage{xcolor}
\usepackage{tablefootnote}
\usepackage{colortbl} 
\usepackage{fancyvrb} 
\usepackage{tcolorbox} 
\def\BibTeX{{\rm B\kern-.05em{\sc i\kern-.025em b}\kern-.08em
    T\kern-.1667em\lower.7ex\hbox{E}\kern-.125emX}}

\definecolor{backcolour}{rgb}{1,1,1} 
\definecolor{codegreen}{rgb}{0,0.6,0}
\definecolor{codegray}{rgb}{0.5,0.5,0.5}
\definecolor{codepurple}{rgb}{0.58,0,0.82}
\definecolor{mymauve}{rgb}{0.58,0,0.82}
\definecolor{mygreen}{rgb}{0,0.6,0}

\lstdefinestyle{mystyle}{
    backgroundcolor=\color{backcolour},   
    commentstyle=\color{codegreen},
    keywordstyle=\color{magenta},
    numberstyle=\tiny\color{codegray},
    stringstyle=\color{codepurple},
    basicstyle=\ttfamily\footnotesize,
    breakatwhitespace=false,         
    breaklines=true,                 
    captionpos=b,                    
    keepspaces=true,                 
    numbers=left,                    
    numbersep=5pt,                  
    showspaces=false,                
    showstringspaces=false,
    showtabs=false,                  
    tabsize=2
}

\lstdefinelanguage{json}{
    basicstyle=\normalfont\ttfamily,
    numbers=left,
    numberstyle=\scriptsize,
    showstringspaces=false,
    breaklines=true,
    backgroundcolor=\color{backcolour},
    stringstyle=\color{codegreen},
    keywordstyle=\color{mymauve},
    commentstyle=\color{mygreen},
    escapeinside={(*@}{@*)},
    morestring=[b]",
    morestring=[d]'
}

\lstdefinestyle{promptstyle}{
    backgroundcolor=\color{backcolour},   
    basicstyle=\ttfamily\footnotesize,
    breakatwhitespace=true,         
    breaklines=true,                 
    captionpos=b,                    
    keepspaces=true,                 
    numbers=none,                    
    numbersep=5pt,                  
    showspaces=false,                
    showstringspaces=false,
    showtabs=false,                  
    tabsize=0
}

\lstdefinelanguage{prompt}{
    basicstyle=\normalfont\ttfamily,
    numbers=none,
    numberstyle=\scriptsize,
    stepnumber=1,
    numbersep=8pt,
    showstringspaces=false,
    breaklines=true,
    breakindent=0pt,
    frame=lines,
    backgroundcolor=\color{backcolour},
    escapeinside={(*@}{@*)},
    morestring=[b]",
    morestring=[d]'
}

\newacro{OSS}[OSS]{Open Source Software}
\newacro{LLM}[LLM]{Large Language Model}

\begin{document}
\acmDOI{10.48550/arXiv.2408.07082}
\pagestyle{plain}
\pagenumbering{arabic}
\title{Evaluating Source Code Quality with Large Language Models: a comparative study}

\author{Igor Regis da Silva Simões}
\email{igor@bb.com.br}
\orcid{0009-0003-8129-3297}
\affiliation{%
  \institution{Universidade de Brasília / Banco do Brasil}
  \city{Brasília}
  \country{Brazil}}

\author{Elaine Venson}
\email{elainevenson@unb.br}
\affiliation{%
  \institution{Universidade de Brasília}
  \city{Brasília}
  \state{DF}
  \country{Brazil}
}

\begin{abstract}
Code quality is an attribute composed of various metrics, such as complexity, readability, testability, interoperability, reusability, and the use of good or bad practices, among others.
Static code analysis tools aim to measure a set of attributes to assess code quality. However, some quality attributes can only be measured by humans in code review activities, readability being an example. Given their natural language text processing capability, we hypothesize that a \ac{LLM} could evaluate the quality of code, including attributes currently not automatable.
This paper aims to describe and analyze the results obtained using \ac{LLM}s as a static analysis tool, evaluating the overall quality of code.
We compared the \ac{LLM} with the results obtained with the SonarQube software and its Maintainability metric for two \ac{OSS} Java projects, one with Maintainability Rating A and the other B. A total of 1,641 classes were analyzed, comparing the results in two versions of models: GPT 3.5 Turbo and GPT 4o.
We demonstrated that the GPT 3.5 Turbo \ac{LLM} has the ability to evaluate code quality, showing a correlation with Sonar's metrics. However,  there are specific aspects that differ in what the \ac{LLM} measures compared to SonarQube. The GPT 4o version did not present the same results, diverging from the previous model and Sonar by assigning a high classification to codes that were assessed as lower quality.
This study demonstrates the potential of \ac{LLM}s in evaluating code quality. However, further research is necessary to investigate limitations such as LLM's cost, variability of outputs and explore quality characteristics not measured by traditional static analysis tools.
\end{abstract}



\captionsetup{font=footnotesize}
\maketitle
\section{Introduction}

Before initiating a modification in software, developer need to read its source code to identify the target sections of the intervention and devise an implementation strategy that maintains coherence with the pre-existing source code. The coding process of the intervention then begins, which should follow a set of best practices. Software developers share practices for writing quality and readable code, also known as \textit{Clean Code} \cite{martin_clean_2009, contieri2023}.

The quality and readability of the source code are factors that impact developer's performance and productivity. A study \cite{cheng_what_2022} with 2,139 developers found that improvements in code quality resulted in an increase in self-declared productivity perception by developers, this being the factor with the highest correlation among the 39 evaluated factors. Code quality is a measurable aspect and monitored through static analysis tools, such as SonarQube\footnote{https://www.sonarsource.com/products/sonarqube/} and code linter tools.

Static analysis tools successfully identify signs of lack of quality in source code. These signs manifest themselves by the occurrences of the so-called \textit{Code Smell} \cite{fowler_refactoring_1999}. This term refers to a class of best practice violations and indicates instances in the code that suggest the existence of a problem requiring the developer's attention. These occurrences can be deterministic, originating from code structures, such as the existence of duplicate code, and are easily identified by existing tools.

\ac{LLM}s are being investigated as a tool to assess code quality, as well as their ability to understand code. As an example, the benchmark dataset \textit{CodeXGLUE (General Language Understanding Evaluation benchmark for CODE)}\cite{lu2021codexglue} evaluates the proficiency of \ac{LLM}s in various tasks and programming languages, such as "Clone detection", "Defect detection", "Code completion", "Code translation", "Code search", "Code repair", "Code summarization" among others \footnote{Microsoft. CodeXGLUE. https://github.com/microsoft/CodeXGLUE}. The characteristics of this benchmark allow analyzing the ability of a model to \textit{understand} code, but there is no specific evaluation regarding a model's ability to \textit{evaluate} the quality of a code.

This work aims to identify the level of ability of \ac{LLM}s in assessing source code quality detectable by tools, as of those that perform static analysis, by answering the questions:
\newcommand{\perguntaDois}{How do the results of an \ac{LLM} compare to the results generated by a static analysis tool?}%
\newcommand{\perguntaTres}{What are the benefits and limitations of using \ac{LLM} for code quality assessment?}%
\begin{itemize}
    \item{\perguntaDois}\label{pergunta2}
    \item{\perguntaTres}\label{pergunta3}
\end{itemize}

To answer these questions, we conducted a quantitative comparative study of the quality rule analysis results from the SonarQube tool with the results obtained by an analysis performed by two versions of the \ac{LLM} ChatGPT. To obtain analysis results similar to those produced by SonarQube, we defined a prompt commanding the \ac{LLM} to issue a \textit{score} ranging from 0 to 100 for the source code being analyzed, considering aspects of readability and overall code quality. The comparison between the quality results produced by SonarQube and the \ac{LLM} was carried out through statistical analyses, seeking to identify the correlation between the generated values. With these results in hand, we analyzed the discrepancies between the \ac{LLM} and SonarQube, and we were able to identify the limitations of the \ac{LLM} compared to SonarQube, as well as the differentials obtained in its use.

The results of this study show that an LLM is capable of capturing aspects related to the overall quality of code, in agreement with what was pointed out by the reference tool. We found that there are characteristics of code quality captured by both the LLM and SonarQube, as well as characteristics captured exclusively by either the LLM or Sonar. We also verified differences in the calculations used by the LLM and SonarQube to assess code quality, raising the divergences of evaluation. Additionally, we observed that the LLM provided a finer analysis compared to the A to E categorization used by Sonar. However, divergent results were noted in different versions of the LLM (GPT 3.5 vs 4o), indicating that the results of this study should not be generalized to all LLMs.

The remainder of this paper is organized into the following sections: Section \ref{sec:barckground} addresses the concepts of static code analysis and \ac{LLM} treated in this study; in Section \ref{sec:relatedwork}, we present an overview of recently published research where \ac{LLM} are applied in Software Engineering activities; following this, Section \ref{sec:methodology} describes the methodology used and Section \ref{sec:results} details the results obtained; finally, Section \ref{sec:conclusion} presents the final considerations, pointing to future work from this study.

\section{Background} \label{sec:barckground}
\subsection{Static Code Analysis}

Static code analysis has become a fundamental part of many software development approaches \cite{thomson_static_2021}. It is present as a strategy to improve code quality and consequently optimize the Code Review process \cite{panichella_would_2015, balachandran_reducing_2013, singh_evaluating_2017}. It is part of steps of modern compilers such as GraalVM \cite{wimmer_graalvm_2021}, being necessary for the generation of optimized native code. It is present in popular tools like Github, which stores hundreds of millions of source code repositories and receives more than 1,000 pushes per minute, from more than 65 million developers. Even with the challenge of meeting this volume, human behavior is pointed out as the main difficulty related to the successful adoption of this type of tool\cite{clem_static_2021}.

The gamification of static code analysis has been studied in order to engage developers, thus addressing the human problem related to its adoption \cite{nguyen_quang_do_gamifying_2018}. In industry-conducted research\cite{distefano_scaling_2019, sadowski_lessons_2018}, it is suggested that success in engaging the developer would be related to bringing static analysis closer to the workflow, providing quick and short feedbacks. It was also identified that the number of defects pointed out by static analyses, which were accepted by developers, would be the success metric to be pursued.

The high level of false positives and false negatives are detractors of this success metric. Added to this is the inability to identify semantic aspects in identifier names, restricting itself to the use of dictionaries and grammatical correction. Examples of these aspects are: Does the code comment assist in understanding the algorithm? Does the name of an identifier (variable, method/function) have relevant meaning to the rest of the purpose of the code, being self-descriptive? Such questions could only be answered by a human reviewer. Limiting the scope of static code analysis tools \cite{gunawardena_concerns_2023}.

\textcite{gunawardena_concerns_2023} identified that 4.6\% of the occurrences pointed out in a code review, refer to the naming of identifiers; 16.4\% refer to API/Javadoc documentation and 4.8\% to comments. Totaling 25.8\% of the points of 417 code reviews analyzed. They also identified that a large portion of these occurrences are not susceptible to automated identification by static analysis tools, requiring a developer's review.

\subsection{Large Language Model}
Large Language Models (\ac{LLM}s) are Artificial Intelligence models, some with tens of billions of parameters, trained using a large volume of data. The success in their development and use is marked by the architecture called \textit{Transformer} \cite{vaswani_attention_2017}, which proposes an attention mechanism. The \ac{LLM}s have demonstrated their superiority over previous models, enabling faster training and better results in benchmark tests. Conventional AI operates on regression and classification problems, analyzing data, classifying and identifying patterns. \ac{LLM}s possess the capabilities of conventional AI, adding what is being called generative AI, which expands this scope of action to create content. Generative AI currently are able to generate text, image, sound and video, allowing \ac{LLM}s to understand commands in text and generate responses in text \cite{zhang_complete_2023}. The use of \ac{LLM} in software engineering activities has increased with the emergence of new research and solutions in addition to its popularization \cite{hou_large_2024}.

\section{Related work} \label{sec:relatedwork}

There is a rich research scenario, with more than 70 different \ac{LLM}s in use, of which 45 follow the architecture called \textit{decoder-only}, just like its most popular example, ChatGPT \cite{hou_large_2024}. \textcite{hou_large_2024} mapped studies related to the use of \ac{LLM}s in Software Engineering, covering the period from 2017 to 2023. When analyzing the distribution of primary articles, they identified that the highest concentration is in the Development activity, with the task of generating code from a natural language prompt comprising the majority of the articles.

The second highest concentration is in Program Repair articles, which study the identification and repair of defects in code. Code completion occupies the third position, popularized by programming assistants like Github Copilot, focused on helping the developer create code. The fourth category is dedicated to the activity of explaining the code, passed as a parameter, in natural language. In the fifth and sixth positions are respectively: The activity of vulnerability detection, in which it is sought to use \ac{LLM}s to detect security weaknesses; and the generation of tests, with emphasis on the largest number of research related to the generation of unit tests.

These tasks suggest the ability of \ac{LLM}'s to understand and analyze codes, but without a dedicated study to evaluate this capability. The task of static analysis is directly related to this capability and has only two articles identified in the study, presented below.

\textcite{hao_ev_2023} present the results obtained in a prototype, using an \ac{LLM} that must execute a pseudocode in order to perform static analysis. The developed prototype receives as a parameter, the pseudocode that is passed as a prompt to the \ac{LLM}, along with information about the task to be performed. The third parameter is the source of the program to be analyzed. The article demonstrates the effectiveness of the proposed technique, comparing its results with the results obtained in the use of the \ac{LLM} without the pseudocode-based technique. There was a 52.94\% improvement in overall accuracy. There was also an improvement in token consumption, the use of the technique presented 46.06\% of the tokens without the use of the technique.

\textcite{mohajer_skipanalyzer_2023} present the results obtained with a static code analysis tool endowed with capabilities of an \ac{LLM}. It has the ability to identify defects through static analysis with \ac{LLM}. It also filters false positives in an additional step with the use of \ac{LLM} and finally performs the correction of the defects found, with modifications also made by \ac{LLM}. The evaluation of the results was done using two common types of defects, \textit{Null Deference} and \textit{Resource Leak}. The defect detection capability was compared with that of the static analysis tool Infer\footnote{Meta. Infer Web Site URL: https://fbinfer.com}. The detection of \textit{Null Deference} was superior to Infer, and the hit rate for \textit{Resource Leak} was also higher. The accuracy in identifying false positives increased significantly for both cases. After analysis with \ac{LLM}, the capability for \textit{Null Deference} improved, and for \textit{Resource Leak} there were no false positives. The automatic correction with \ac{LLM} presented high correctness for \textit{Null Deference} and \textit{Resource Leak}, with almost all the generated code being syntactically correct.

Both articles demonstrate promising results with the use of \ac{LLM} for static code analysis. However, the focus of both is on defect detection. The ability of \ac{LLM}s in static analysis aimed at overall code quality, which is the objective of this study, was not evaluated.

The systematic mapping of \textcite{hou_large_2024} also identified seven articles classified as related to the Code Review task of the Maintenance activity. Among them, two articles deal with performing program editing based on Code Review comments \cite{zhang_coditt5_2022} and evaluating the readability and explainability of programs generated by \ac{LLM}'s \cite{liu_reliability_2024}.

Five other studies, \cite{guo_exploring_2024, li_auger_2022, lu_llama-reviewer_2023, tufano_using_2022, sghaier_multi-step_2023}, are dedicated to automating Code Review. They use \ac{LLM} to identify points of improvement in the code and generate review information.

\textcite{sghaier_multi-step_2023} evaluated the effectiveness of \ac{LLM}'s, pre-trained and fine-tuned, in predicting reviews from the source code.
\textcite{tufano_using_2022} uses a Text-To-Text Transfer Transformer (T5) model to perform reviews recommending improvements to the code and perform the proposed changes.
\textcite{li_auger_2022} use the T5 model from the work of \textcite{tufano_using_2022}, fine-tuned for Java code, as part of a framework (AUGER) presented in the study, comparing it to the LSTM, CopyNet and CodeBERT models.
\textcite{lu_llama-reviewer_2023} use a LLaMA model in the composition of their framework to automate the Code Review process, comparing it to seven other \ac{LLM}s.

Four of these works, \cite{li_auger_2022, lu_llama-reviewer_2023, tufano_using_2022, sghaier_multi-step_2023}, performed the pre-training and fine-tune of the models and compared the results to other models, used as baselines. Code Review datasets were also used to execute the experiments.

The fifth study, conducted by \textcite{guo_exploring_2024}, is the only one that evaluates a pre-trained model without performing fine-tune. The chosen model is ChatGPT, in its GPT-3.5-turbo and GPT-4 versions. A reference model, specialized in code review, was used, using a code review dataset. To mitigate the risk that the GPT models have used the public datasets in their training, an additional dataset was used. Five prompt strategies were evaluated: (P1) simplest prompt, (P2) P1 + scenario description, (P3) P1 + detailed requirements, (P4) P1 + concise requirements and (P5) P4 + scenario description. Five temperature levels were also evaluated: 0, 0.5, 1.0, 1.5 and 2.0. The best results were obtained with the combination of temperature = 0 and prompt P2. The P2 prompt used the prompt pattern named by \textcite{white_prompt_2023} as Persona. The performance of the ChatGPT models in performing code refinements based on review comments was evaluated, similar to the work of \textcite{zhang_coditt5_2022}. Its ability to analyze the quality of the code that underwent review, which is the focus of this study, was not evaluated. The results obtained were promising, but still limited.

\section{Methodology} \label{sec:methodology}

\newcommand{\itemA}{Project 1 Analysis - Quarkus}%
\newcommand{\itemAUm}{Selection of excerpts for analysis}%
\newcommand{\itemADois}{Sonar Analysis}%
\newcommand{\itemATres}{\ac{LLM} Analysis}%
\newcommand{\itemAQuatro}{Results comparison}%

\newcommand{\itemB}{Project 2 Analysis - Shattered Pixel Dungeon}%
\newcommand{\itemBUm}{Selection of excerpts for analysis}%
\newcommand{\itemBDois}{Sonar Analysis}%
\newcommand{\itemBTres}{\ac{LLM} Analysis}%
\newcommand{\itemBQuatro}{Results comparison}%

\newcommand{\itemC}{Analysis with second \ac{LLM}}%
\newcommand{\itemCUm}{Quarkus Project}%
\newcommand{\itemCDois}{SPDungeon Project}%

This observational study compares code quality metrics generated from the use of \ac{LLM}s with metrics generated by the static analysis tool SonarQube.

This section describes the procedures used for: (1) selection of projects and classes to be analyzed; (2) selection of metrics used for comparison; (3) definition of the prompt to extract the metrics from the \ac{LLM}; and (4) comparison of results.

\subsection{Selection of Projects and Classes}

The SonarCloud site\footnote{SonarCloud https://sonarcloud.io/explore/projects} provides access to public analyses for \ac{OSS} projects. At the time of this research, the language with the highest number of analyzed projects available was Java (excluding HTML). Java is also the most studied language for source code metrics \cite{nunez-varela_source_2017} and code comprehensibility \cite{wyrich_40_2024}. Two projects with a large volume of source code lines were chosen, one project with a maintainability rating of A and another with rating B, according to Sonar. Projects with an A rating have a technical debt cost for remediation equal to or less than 5\% of the total software cost. For remediation costs between 6\% and 10\%, the project receives a B rating. The same criterion applies to each Java class file.

The first filter of the tool was applied, for projects with more than 500 Kloc, returning 86 projects. A second criterion was applied, selecting only renowned \ac{OSS} projects (with more than 10k stars), leaving the following projects: Apache Hadoop, Druid, Spark, Flink, ElasticSearch and Quarkus. A third condition was applied, selecting projects with analyses carried out in the 10 days prior to this work and no divergence of activity date of the repository on Github and the last analysis of Sonar, leaving only the Quarkus. No project with rating B was found among the 86. We expanded the filter to the second interval allowed by the tool (100 - 500 Kloc), returning 978 projects. Seven projects were found on SonarCloud with a B maintainability rating, of which five were excluded because they did not have an analysis carried out in 2024 or because they had a divergence of activity date of the repository on Github and the last analysis of Sonar, which would lead the \ac{LLM} to analyze a different source from Sonar. And finally, one project was excluded for not having a source available on Github, leaving only one project with a B rating.

The selected projects were Quarkus\footnote{Quarkus https://github.com/quarkusio/quarkus}, a Cloud Native, (Linux) Container First framework for writing Java applications, and Shattered Pixel Dungeon (SPDungeon)\footnote{SPDungeon https://github.com/ismvru/shattered-pixel-dungeon}, an open-source traditional roguelike dungeon crawler with randomized levels and enemies, and hundreds of items to collect and use. It's based on the source code of Pixel Dungeon, by Watabou. Table \ref{tab:projects} presents the main characteristics of the two projects.

\begin{table}[ht]
\centering
\footnotesize
\caption{Characteristics of the selected projects}
\label{tab:projects}
    \begin{tabular}{lrr}
        \toprule
        \textbf{Characteristic} & \textbf{Quarkus} & \textbf{SPDungeon} \\
        \midrule
        Sonar Rating & A & B \\
        Github stars & 13.4K & 4.5K \\
        Releases & 319 & 47 \\
        Lines of code & 725 KLoc & 134 KLoc \\
        Total of classes & 11.493 & 1.790 \\
        Analyzed classes & 644 & 997 \\
        \bottomrule
    \end{tabular}
\end{table}
The analyses carried out on SonarQube for the SPDungeon project are from a Github fork, with the same sources as the original repository.

The analysis was conducted for the project classes with the highest possible number of lines of code within the \ac{LLM} token limit, resulting in files with fewer than 800 lines of source code. Automated test classes and automatically generated classes were disregarded. To preserve the state of the sample used in this study, the source code of each class was copied to a project on Github 
\makeatletter
\if@ACM@anonymous
    \textit{(link hidden for anonymization purposes)}.
\else
    \cite{repoGitHubIgorBB}.
\fi
\makeatother 

\subsection{Analyzed Metrics}

SonarQube offers a range of code metrics resulting from its analysis. The following metrics were selected for comparison (in parentheses is indicated how they will be referenced in this article): 
\begin{enumerate}
    \item Number of Code Smells of a class (Code Smells).
    \item Comment Lines Density (\% Comments).
    \item Cognitive complexity (Cog. Complexity).
    \item Complexity (Complexity).
    \item Lines (Lines) representing the amount of code.
    \item Statement (Statement), number of statements of a class. 
\end{enumerate}
Those metrics are some of the most used in researches related to code source code metrics \cite{nunez-varela_source_2017} also measured by SonarQube. The exceptions are Code Smells being a composition of many metrics and Cognitive Complexity as a new metrics introduced into the tool \footnote{SonarQube Cognitive Complexity White Paper https://www.sonarsource.com/docs/CognitiveComplexity.pdf} as a complementary metric to Cyclomatic Complexity with a enfasis to code understandability \cite{campbell_cognitive_2018}.

\subsection{\ac{LLM}'s Prompt Definition}

The primary \ac{LLM} used for this study was ChatGPT 3.5 Turbo. This version presents a token cost 60 times lower than ChatGPT 4 and 10 times lower than the latest version, ChatGPT 4o. \textcite{laskar_systematic_2023} identified that newer versions of ChatGPT do not guarantee better results. The aim of this study was not to compare different \ac{LLM}s, therefore the version with the lower cost was primarily used. As previous research found that newer versions may present worse results, an evaluation was also carried out with the latest version of ChatGPT, the 4o.

The analysis carried out with the \ac{LLM} requires the elaboration of a suitable prompt for this purpose, the prompt of this study was based on specific patterns for ChatGPT \cite{white_prompt_2023} and the lessons learned by \textcite{guo_exploring_2024}. The first consists of guiding the \ac{LLM} to assume a persona capable of performing the analysis with a certain knowledge. For this purpose, we elaborated the passage \textit{"The assistant is a seasoned senior software engineer, with deep Java Language expertise, doing source code evaluation as part of a due diligence process, these source code are presented in the form of a Java Class File. Your task is to emit a score from 0 to 100 based on the readability level and overall quality of the source code presented."}. 

To allow the evaluation of the score assigned by the \ac{LLM}, we requested that it present the explanation for the grade by passing guidelines on its format \textit{"- The ”explanation” attribute must not surpass 450 characters and MUST NOT contain special characters or new lines."}. This explanation have an auditing purpose only, by having the LLM to elaborate over the reasons of the generated score. To facilitate the reading of the \ac{LLM}'s responses and even its processing via software, we complemented it with the approach of defining a response template that the \ac{LLM} must follow \textit{"Your answers MUST be presented ONLY in the following json format: {”score”:”NN\%”, reasoning:”your explanation about the score”}"}.

We used three patterns mapped by \textcite{white_prompt_2023} (\textit{Persona, Template and Reflection}) to elaborate the prompt that guides the \ac{LLM} on how to respond to the task assigned to it. The elaborated passages composed the command of the \ac{LLM} as System Prompt:

\begin{tcolorbox}[colback=gray!10, colframe=gray!80, title=System Prompt]
\begin{Verbatim}[fontsize=\scriptsize]
The assistant is a seasoned senior software engineer, with deep Java
Language expertise, doing source code evaluation as part of a due 
diligence process, these sourcode are presented in the form of a Java
Class File. Your task is to emit a score from 0 to 100 based on the 
readability level and overall quality of the sourcode presented. 
Your answers MUST be presented ONLY in the following json format: 
\{"score":"NN\%", reasoning:"your explanation about the score"\}
- The "explanation" attribute must not surpass 450 characters and 
MUST NOT contain especial characters or new lines.
\end{Verbatim}
\end{tcolorbox}
\label{prompt}

In addition to the System Prompt, the User Prompt needs to be developed, a section in which the command that the \ac{LLM} is expected to execute is given, and the code to be analyzed is inserted:

\begin{tcolorbox}[colback=gray!10, colframe=gray!80, title=User Prompt]
\begin{Verbatim}[fontsize=\scriptsize]
Evaluate the following Java source code: SOURCE-CODE This is the end of 
the class file, the assistant should present your json answer:
\end{Verbatim}
\end{tcolorbox}
\label{prompt}

\paragraph{Other parameters:} According to the OpenAI documentation\footnote{OpenAI Documentation - https://platform.openai.com/docs/guides/text-generation/how-should-i-set-the-temperature-parameter}, the \textit{Temperature} parameter regulates the degree of consistence of the \ac{LLM} with reference to the command given to it. The value is a range between 0 and 2, where 0 regulates the \ac{LLM} to offer factual responses and 2 regulates it to a more creative mode in which the \ac{LLM} can generate responses without factual bases. Thus, for the task of code analysis, it is expected that the responses presented have a factual basis strictly related to the presented code, setting this parameter to the value 0 \textit{(zero)}. \textcite{guo_exploring_2024} identified that this value generates the best results when performing prompts that seek to analyze code. 

The context window of an LLM is determined by the combined length of its input and output. To maximize the number of lines of code into the input and avoid the truncation of the output, we limited the LLM response to 120 tokens, leaving the remaining tokens for the input (prompt + source code). A LLM can't count characters but solely tokens, the 120 token limit used as output parameter guarantee the generated response to be near to 450 characters \cite{shin_large_2024}.

\subsection{Comparative Analysis}

The results between the values of the metrics collected via Sonar and via \ac{LLM} for the two projects were compared using Spearman correlation algorithms. Attributes such as mean, mode, standard deviation, median, and percentage of outliers were also measured to allow a better understanding of the data distribution. A listing of the distribution of ratings assigned by Sonar and scores generated by the \ac{LLM} was made. This listing was cross-referenced to find discrepancies between both approaches. The discrepancies were analyzed and discussed.

\section{Results} \label{sec:results}

In this section, we present the result of the code quality analysis for each of the two selected \ac{OSS} projects. For each of them, we present the characteristics of the analyzed classes, the result of the analysis from Sonar, the result of the analysis from the \ac{LLM}s, and finally, the comparison of the results.

\begin{table}[ht]
\footnotesize
    \begin{minipage}{.4\linewidth}
    \centering
        \caption{Rating SonarQube}
        \label{tab:tabelaManutenibilidadeAmostraQuarkus}    
        \begin{tabular}{cr}
            \toprule
            \textbf{Rating} & \textbf{Classes} \\
            \midrule
            A & 615 \\
            B & 21 \\
            C & 6 \\
            D & 2 \\
            E & 0 \\
            \bottomrule
        \end{tabular}
    \end{minipage}
    \begin{minipage}{.4\linewidth}
    \centering
        \caption{Score LLM}
        \label{tab:tabelaScoreLLM1}    
        \begin{tabular}{cr}
            \toprule
            \textbf{Score} & \textbf{Classes} \\
            \midrule
            90 & 176 \\
            85 & 421 \\
            80 & 30 \\
            75 & 10 \\
            70 & 4 \\
            60 & 1 \\
            50 & 2 \\
            \bottomrule
        \end{tabular}
    \end{minipage}
\end{table}

\subsection{\itemA}
\subsubsection{\itemAUm}
Classes selected from the core layer of the source code of the Java project called Quarkus will be used, which already has analyses performed on the public site of SonarQube \footnote{Quarkus https://sonarcloud.io/summary/overall?id=quarkusio\_quarkus}.

\subsubsection{\itemADois} The core layer of Quarkus has 644 classes that fit the selection criteria for this work, totaling 55,839 lines of code and 1,626 Code Smells. The distribution of the maintainability rating of the sample, presented in Table \ref{tab:tabelaManutenibilidadeAmostraQuarkus}, indicates that the vast majority of the sample classes have a high maintainability index, with 95.49\% of the classes in rating A and 3.26\% in rating B, according to the metrics applied by Sonar.

Table \ref{tab:tabelaEstatisticasSonar} presents the summary statistics for the metrics collected from Sonar. With the exception of the percentage of comments, there is a high number of outliers in all attributes. Analyzing the statistical data of these attributes, it is noticed that most of the sample classes have no Code Smells, low complexity (both metrics), few lines of code, comments, and statements.

\begin{table}[ht]
\footnotesize
\caption{Sonar's metrics statistics}
\label{tab:tabelaEstatisticasSonar}
    \begin{tabular}{lrrrrrr}
        \toprule
        \textbf{Atributo} & \textbf{Média} & \textbf{Std.Dev.} & \textbf{Mediana} &\textbf{Moda} & \textbf{Outliers} \\
        \midrule
        Code Smells & 2.52 & 5.14 & 0 & 0 & 9.78\%\\
        \% Comments & 13.92 & 16.37 & 7.7 & 0 & 4.35\%\\
        Cog. Complexity & 10.16 & 25.49 & 0 & 0 & 15.22\%\\
        Complexity & 11.53 & 19.81 & 4 & 0 & 12.27\%\\
        Lines & 86.70 & 116.64 & 42.5 & 18 & 10.71\%\\
        Statements & 24.25 & 47.87 & 5 & 0 & 12.73\%\\
        \bottomrule
    \end{tabular}
\end{table}

\subsubsection{\itemATres}
The number of classes for each score assigned by the \ac{LLM} is listed in Table \ref{tab:tabelaScoreLLM1}. There are 421 occurrences of score 85, representing 65.37\% of the total, and 176 occurrences with score 90, accounting for 27.32\%. A strong concentration of evaluations around score 85 is observed, with 97.36\% of the classes having a score of 80 or higher.

Unlike the attributes derived from Sonar, for the score values produced by the \ac{LLM}, there is a low occurrence of outliers, only 2.64\% (score less than or equal to 70).

\subsubsection{\itemAQuatro}

\begin{figure}[ht]
  \centering
  \includegraphics[width=0.5\columnwidth]{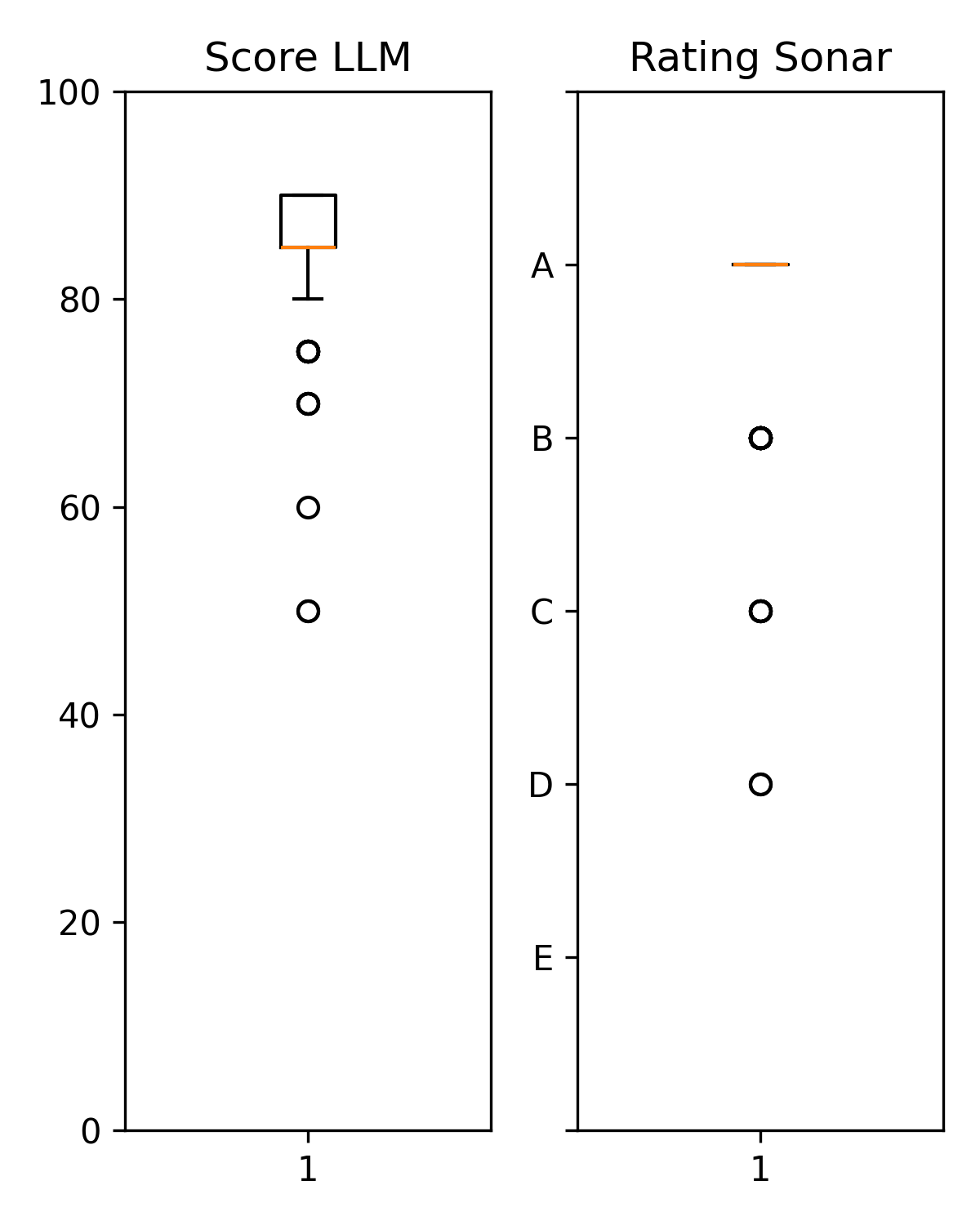}
  \vspace*{-10pt}
  \caption{LLM scores vs Sonar rating}
  \label{fig:boxScoreAndRatingSonarQuarkus} 
\end{figure}

The SonarQube analysis shows that the group of classes with rating A corresponds to 95.49\% of the sample, while the analysis with \ac{LLM} shows that the group of classes with a score of 85 or higher corresponds to 92.70\%, or 97.36\% for a score of 80 or higher. The Rating assigned by SonarQube is a categorical variable that generated a concentration with value A. Figure \ref{fig:boxScoreAndRatingSonarQuarkus} presents the apparent correlation between the distribution of the Sonar rating and the score assigned by the LLM.

\begin{table}[ht]
    \centering
    \small
    \caption{Distribution of SonarQube vs LLM (Quarkus) reviews}
    \label{tab:distribuicaoQuarkusSonarScore35}
        \begin{tabular}{crrrr}
            \toprule
            \scriptsize{\textbf{LLM}} & \multicolumn{4}{c}{\scriptsize{\textbf{Sonar Rating}}} \\
            \cmidrule{2-5}
            \scriptsize{\textbf{Score}} &  A &  B & C & D \\
            \hline
            50 &    0 &   \cellcolor{yellow!25} 2 &    0 &    0 \\
            60 &    1 &    0 &    0 &    0 \\
            70 &    4 &    0 &    0 &    0 \\
            75 &    9 &    1 &    0 &    0 \\
            80 &   27 &    1 &    2 &    0 \\
            85 &  404 &   13 &    2 &  \cellcolor{yellow!25}  2 \\
            90 &  170 &    4 &    2 &    0 \\
            \bottomrule
        \end{tabular}
\end{table}

Table \ref{tab:distribuicaoQuarkusSonarScore35} presents the relationship between the scores assigned by the \ac{LLM} and the rating assigned by Sonar at the class level. It can be observed that there are discrepancies between the \ac{LLM} and SonarQube, as exemplified by the cells highlighted in yellow. Two classes with a B rating received a score of 50. Both classes are empty, without attributes, methods, or comments, as can be observed in Listings\footnote{We removed the package and import declarations} \ref{lst:ClassesVaziasConstPoolPredicate} and \ref{lst:ClassesVaziasDescription}. In Listing \ref{lst:AvaliacaoLLMClassesVazias}, we have the score and the justification provided by the \ac{LLM} for the assigned grade. It is notable that the absence of code and documentation determined the score assigned by the LLM.

Upon detailed analysis of the SonarQube evaluation, we found the existence of only one Code Smell of the \textit{Maintenability} type for each class. It is a rule violation, considered of the \textit{Minor} type by SonarQube. This rule warns about the use of empty classes in Java, considered a bad practice reducing code maintainability by generating comprehension confusion. The weight of the violation is given by SonarQube due to the estimated effort for the correction of the pointing, which for the case of this rule, is five minutes of work.

\begin{lstlisting}[style=mystyle, language=java, firstnumber=1, basicstyle=\tiny, caption={Classe ConstPoolPredicate (score 50)}, label={lst:ClassesVaziasConstPoolPredicate}]
public class ConstPoolPredicate {
}
\end{lstlisting}
\begin{lstlisting}[style=mystyle, language=java, firstnumber=1, basicstyle=\tiny, caption={Classe Description (score 50)}, label={lst:ClassesVaziasDescription}]
public class Description {
}
\end{lstlisting}

\begin{lstlisting}[style=mystyle, language=json, firstnumber=1, basicstyle=\scriptsize, caption={Samples of LLM's code reviews}, label={lst:AvaliacaoLLMClassesVazias}]
ConstPoolPredicate: {"score":"50","reasoning":"The code is very short and lacks any functionality or context. It is difficult to evaluate the quality of the code based on this class alone. However, the code follows standard Java naming conventions and is properly formatted."}
Description: {"score":"50","reasoning":"The code is very short and doesn't contain any logic. It's hard to evaluate the readability and quality of the code based on just one empty class."}
\end{lstlisting}

We also analyzed the two classes with a D rating that received a high score (85). In Listings \ref{lst:QuarkusBindException} and \ref{lst:ClasseSnapStartRecorder}, we have the source of the classes. Unlike the \textit{ConstPoolPredicate} and \textit{Description} classes, these contain code and comments. The \textit{SnapStartRecorder} class is short, but its content did not receive a negative evaluation from the LLM. In the justifications presented in Listing \ref{lst:AvaliacaoLLMClassesRatingAlto}, we note a positive reference to the code structure, correct use of Java language capabilities, and a recommendation for documentation improvement, besides a comment about the good variable naming. The \textit{QuarkusBindException} class received similar comments.

The \textit{QuarkusBindException} class has only one violation, considered \textit{Major}, with an estimated effort of four hours of work for correction. The \textit{SnapStartRecorder} class has six violations of three SonarQube rules, totaling an estimated effort of one hour and 40 minutes. All violations compromise maintainability, two of the rules have a related CWE\footnote{Rule https://rules.sonarsource.com/java/RSPEC-1104/} \footnote{Rule https://rules.sonarsource.com/java/RSPEC-1444/}, the other refers to the risk of problems in multi-threading scenarios\footnote{Rule https://rules.sonarsource.com/java/RSPEC-2696/}. None of the rules were pointed out by the LLM.

We conclude that the evaluation carried out by the \ac{LLM} did not take into account the same aspects as SonarQube.

\begin{lstlisting}[style=mystyle, language=java, firstnumber=1, basicstyle=\tiny, caption={Classe QuarkusBindException (score 85)}, label={lst:QuarkusBindException}]
/**
 * An exception that is meant to stand in for {@link BindException} and provide information
 * about the target which caused the bind exception.
 */
public class QuarkusBindException extends BindException {
    private final List<Integer> ports;

    private static String createMessage(List<Integer> ports) {
        return "Port(s) already bound: " + ports.stream().map(i -> Integer.toString(i)).collect(Collectors.joining(", "));
    }

    private static void assertPortsNotEmpty(List<Integer> ports) {
        if (ports.isEmpty()) {
            throw new IllegalStateException("ports must not be empty");
        }
    }

    public QuarkusBindException(Integer... ports) {
        this(Arrays.asList(ports));
    }

    public QuarkusBindException(List<Integer> ports) {
        super(createMessage(ports));
        assertPortsNotEmpty(ports);
        this.ports = ports;
    }

    public QuarkusBindException(BindException e, Integer... ports) {
        this(e, Arrays.asList(ports));
    }

    public QuarkusBindException(BindException e, List<Integer> ports) {
        super(createMessage(ports) + ": " + e.getMessage());
        assertPortsNotEmpty(ports);
        this.ports = ports;
    }

    public List<Integer> getPorts() {
        return ports;
    }
}
\end{lstlisting}
\begin{lstlisting}[style=mystyle, language=java, firstnumber=1, basicstyle=\tiny, caption={Classe SnapStartRecorder (score 85)}, label={lst:ClasseSnapStartRecorder}]
/**
 * Stores the optimizations to execute for SnapStart.
 */
@Recorder
public class SnapStartRecorder {
    public static boolean enabled = false;
    public static boolean fullWarmup = false;

    public void register(boolean fw) {
        enabled = true;
        fullWarmup = fw;
    }
}
\end{lstlisting}

\begin{lstlisting}[style=mystyle, language=json, firstnumber=1, basicstyle=\scriptsize, caption={Samples of LLM's code reviews}, label={lst:AvaliacaoLLMClassesRatingAlto}]
QuarkusBindException: {"score":"85","reasoning":"The code is well-structured and follows Java conventions. The use of streams and lambdas is appropriate. However, the code could benefit from more comments and documentation to explain the purpose of the class and its methods."}
SnapStartRecorder: {"score":"85","reasoning":"The code is well-structured and easy to read. The variable names are clear and concise. However, the purpose of the class and its methods could be better documented."
\end{lstlisting}

We used Spearman’s correlation to test the correlation between the Sonar attributes and the score value of the \ac{LLM} (Table \ref{tab:tabelaCorrelacaoSpearman1}). Spearman’s correlation was used given the non-normal distribution of the data. Through this calculation, we noticed a moderate inverse correlation between the Code Smells, Cog Complexity, Complexity, Lines and Statements in relation to the LLM score. Higher values for those attributes correlate to lower levels of code quality and readability. This inverse correlation to LLM score confirms that the LLM score is positively correlated to source code quality. The \% Comments showed a weak positive correlation and a level of statistical relevance that, although sufficient, is much lower than that presented in the other attributes.

\begin{table}[ht] 
\centering
\footnotesize
\caption{Spearman correlation for SonarQube metrics}
\label{tab:tabelaCorrelacaoSpearman1}
\begin{tabular}{lrr}
    \toprule
    \textbf{Atributo} & \textbf{Correlação} & \textbf{Valor-p} \\
    \midrule
    Code Smells & -0.403 & 1.269e-26\\
    \% Comments & 0.129 & 0.001\\
    Cog. Complexity & -0.361 & 2.667e-21\\
    Complexity & -0.330 & 6.785e-18\\
    Lines & -0.390 & 7.880e-25\\
    Statements & -0.356 & 9.286e-21\\
    \bottomrule
\end{tabular}
\end{table}

\subsection{\itemB}

\subsubsection{\itemBUm}Selected classes from the core layer of the project source code will be used, as was done for Project 1. The most recent analysis on the public SonarQube site \footnote{Shattered Pixel Dungeon https://sonarcloud.io/project/overview?id=ismvru\_shattered-pixel-dungeon} was carried out on 07/02/2024, which is the same date as the last activity of the repository on Github, thus ensuring symmetry between the sources used for the analysis with \ac{LLM} and the Sonar data. The criterion used for the selection of classes and use of SonarQube attributes was the same as for project 1, the classes used in the analysis were also preserved.

\subsubsection{\itemBDois}

\begin{table}[ht]
    \footnotesize
    \begin{minipage}{0.4\columnwidth}
        \caption{Rating SonarQube}
        \label{tab:tabelaManutenibilidadeAmostraShattered}
        \centering
        \begin{tabular}{cr}
            \toprule
            \textbf{Rating} & \textbf{Classes} \\
            \midrule
            A & 670 \\
            B & 109 \\
            C & 86 \\
            D & 103 \\
            E & 29 \\
            \bottomrule
        \end{tabular}
    \end{minipage}
    \begin{minipage}{0.4\columnwidth}
        \centering
        \caption{Score LLM}
        \label{tab:tabelaScoreLLM2}        
        \begin{tabular}{cr}
            \toprule
            \textbf{Score} & \textbf{Classes} \\
            \midrule
            90 & 1 \\
            85 & 107 \\
            80 & 301 \\
            75 & 467 \\
            70 & 117 \\
            60 & 3 \\
            50 & 1 \\
            \bottomrule
        \end{tabular}
    \end{minipage}
\end{table}
    
The core layer of SPDungeon contains 997 classes that fit the selection criteria for this work, totaling 148,701 lines of code and 4,432 Code Smells. Table \ref{tab:tabelaManutenibilidadeAmostraShattered} presents the distribution of the maintainability rating of the sample. Unlike the previous project, which had a concentration of 95.49\% of classes with an A rating, this project presents a 67.2\% concentration in this layer, in addition to presenting a higher occurrence of classes in the lower quality ratings.

\begin{table}[ht]
\footnotesize
\caption{Sonar’s metrics statistics}
\label{tab:tabelaEstatisticasSonarShattered}
    \begin{tabular}{lrrrrrr}
        \toprule
        \textbf{Atributo} & \textbf{Média} & \textbf{Std.Dev.} & \textbf{Mediana} &\textbf{Moda} & \textbf{Outliers}\\
        \midrule
        Code Smells & 4.54 & 7.48  & 2 & 1 & 8.83\%\\
        \% Comments & 2.34 & 3.39 & 1.2 & 0 & 4.81\%\\
        Cog. Complexity & 21.68 & 35.51 & 8 & 0 & 10.13\%\\
        Complexity & 22.04 & 28.71 & 11 & 4 & 9.73\%\\
        Lines & 149.14 & 131.72 & 101 & 51 & 9.83\%\\
        Statements & 48.48 & 59.53 & 26 & 9 & 10.23\%\\
        \bottomrule
    \end{tabular}
\end{table}

Despite presenting a lower percentage of outliers than Project 1, there is still a high outliers occurrence, with the exception of the percentage of comments again, as per the data in Table \ref{tab:tabelaEstatisticasSonarShattered}. Analyzing the statistical data of these attributes, it is noticed that most of the classes in the sample have at least one Code Smell, a complexity per class double the value of Project 1 (both metrics), almost double the lines of code, even fewer comments, and double the statements.

\subsubsection{\itemBTres}
The number of classes for each score assigned by the \ac{LLM} is listed in Table \ref{tab:tabelaScoreLLM2}. There are 108 occurrences of score 85, representing 10.73\%, and one occurrence with score 90 representing 0.1\% compared to 27.32\% of the previous project. There is a strong concentration of evaluations around score 75 (46.84\%), with 41.02\% of the classes having a score of 80 or higher. The occurrence of outliers was even lower for Project 2, only 0.50\%, but the outliers have scores of 90, 60, and 50, both sides of the scale.
\subsubsection{\itemBQuatro}

\begin{figure}[ht]
  \centering
  \includegraphics[width=0.5\columnwidth]{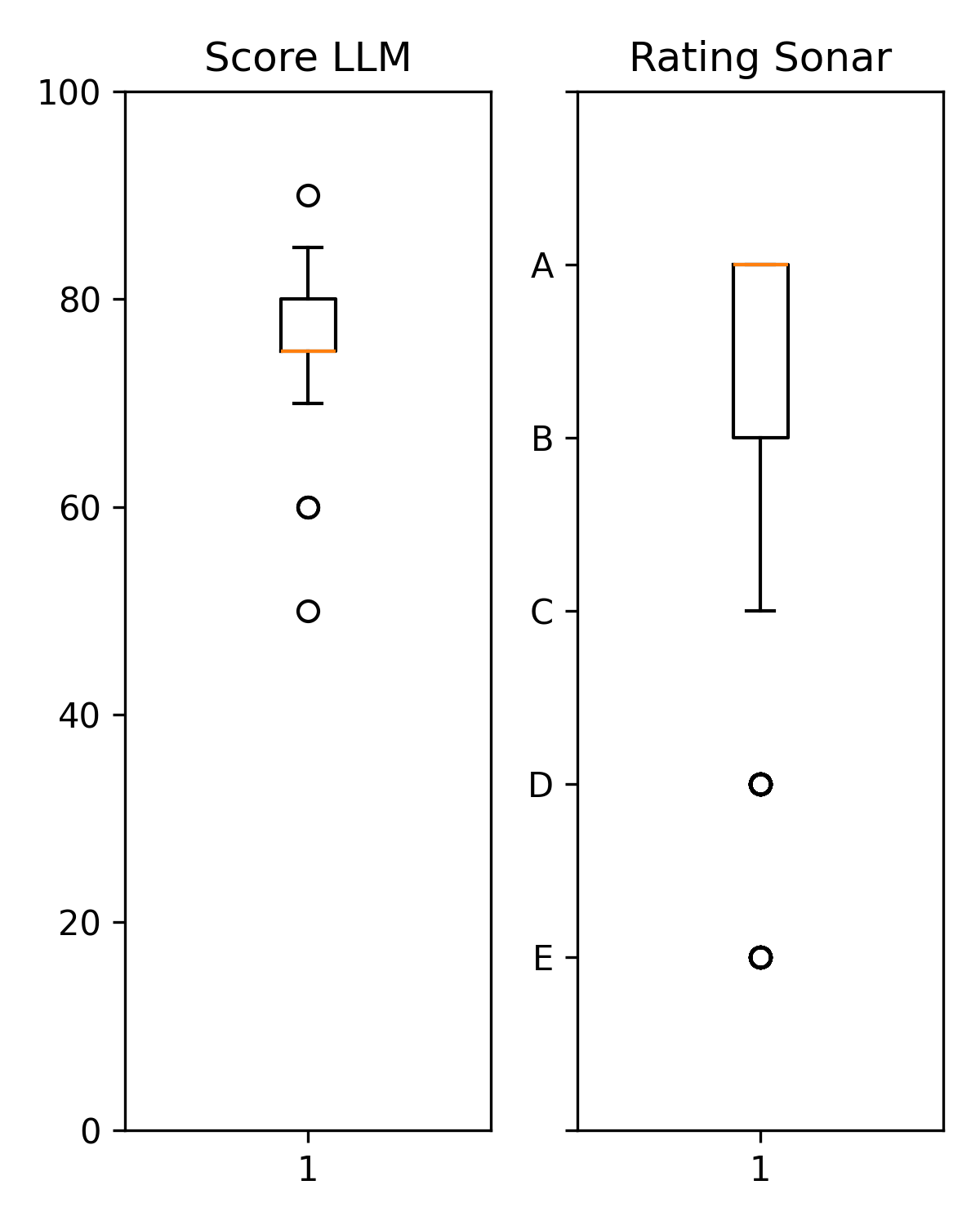}
  \vspace*{-10pt}
  \caption{LLM score vs Sonar rating} 
  \label{fig:boxScoreAndRatingSonarShattered}
\end{figure}

The SonarQube analysis shows that the group of classes with an A rating corresponds to 67.2\%, and another 29.88\% are distributed among the B, C, and D ratings, while the analysis with \ac{LLM} shows that the group of classes with a score of 80 or higher corresponds to 41.02\%, and the scores between 70 and 75 correspond to 58.57\%. For Project 2, we have a larger distribution of classes among various scores by the LLM. As in the Sonar analysis, in both cases there is a tendency to represent a lower code quality when compared to Project 1. In Figure \ref{fig:boxScoreAndRatingSonarShattered}, this approximation of both metrics towards the center is visually noticeable.

\begin{table}[ht]
    \centering
    \small
    \caption{Distribution of Sonar vs LLM (SP Dungeon) reviews}
    \label{tab:distribuicaoSPDungeonSonarScore35}
    \begin{tabular}{crrrrr}
        \toprule
        \scriptsize{\textbf{LLM}} & \multicolumn{5}{c}{\scriptsize{\textbf{Sonar Rating}}} \\
        \cmidrule{2-6}
        \scriptsize{\textbf{Score}} &  A &  B &  C &  D & E \\
        50 &    0 &   \cellcolor{yellow!25} 1 &    0 &    0 & 0 \\
        60 &   \cellcolor{yellow!25} 2 &   \cellcolor{yellow!25} 1 &    0 &    0 & 0 \\
        70 &   78 &   20 &   10 &    7 & 2 \\
        75 &  350 &   39 &   39 &   36 & 3 \\
        80 &  173 &   44 &   22 &   40 & 22 \\
        85 &   67 &    4 &   15 &   20 & \cellcolor{yellow!25} 1 \\
        90 &    0 &    0 &    0 &    0 & \cellcolor{yellow!25} 1 \\
        \bottomrule
    \end{tabular}
\end{table}
Analyzing the relationship of the \ac{LLM} scores with each Sonar Rating, in Table \ref{tab:distribuicaoSPDungeonSonarScore35}, we note that even classes with a high rating (A and B) received low scores from the LLM. Two classes with a B rating received scores of 50 and 60 respectively, and two other classes with an A rating received a score of 60. No class with a C, D, or E rating received a score below 70, contrary to expectations. In addition, two classes with an E rating received scores of 85 and 90 respectively.

In Listing \ref{lst:ClassesRatingABScoreBaixoSPD}, we have the score and respective justifications provided by the \ac{LLM} for the 4 classes with low scores. We note references to problems with code organization, indentation, and documentation, as well as some Code Smells. Some of the aspects pointed out are captured by SonarQube, such as the use of magic numbers or complex methods (cyclomatic complexity). Others are not identified by Sonar, such as methods that could receive better names, poorly formatted code, and appropriate documentation (SonarQube restricts itself to counting the percentage of documentation). The classes with these scores are: \textit{SurfaceScene}, \textit{Combo}, \textit{ElementalStrike}, and \textit{ShadowClone}, having between 430 and 580 lines of code. Because of its size, they will not be listed in the article, but can be consulted in the project related to this research, on GitHub
\makeatletter
\if@ACM@anonymous
    \textit{(link hidden for anonymization purposes)}.
\else
    \cite{repoGitHubIgorBB}.
\fi
\makeatother

\begin{lstlisting}[style=mystyle, language=json, firstnumber=1, basicstyle=\scriptsize, caption={Samples of LLM's code reviews}, label={lst:ClassesRatingABScoreBaixoSPD}]
SurfaceScene: {"score":"50","reasoning":"The code is poorly formatted and lacks comments, making it hard to read and understand. There are also some unused imports and variables, and some methods could be better named. However, the code seems to be functional and follows some good practices, such as using constants and separating concerns into different classes."}
Combo: {"score":"60","reasoning":"The code has good use of comments and follows a standard license. However, it lacks proper indentation and has some long methods and classes, which makes it hard to read and maintain. There are also some unused imports and some code smells, such as the use of public fields and the lack of encapsulation in some methods."}
ElementalStrike: {"score":"60","reasoning":"The code is well-organized and follows good naming conventions. However, there are some issues with readability due to the use of long method chains and nested if statements. Additionally, there are some areas where the code could be simplified and made more concise."}
ShadowClone: {"score":"60","reasoning":"The code has some good practices, such as the use of comments and proper indentation. However, there are some issues, such as the lack of proper documentation, the use of wildcard imports, and the presence of some long methods and classes. Additionally, there are some code smells, such as the use of magic numbers and the mixing of concerns in some methods."}
\end{lstlisting}

\begin{table}[ht]
    \small
    \centering
    \caption{Violations of SonarQube rules for classes on Listing \ref{lst:ClassesRatingABScoreBaixoSPD}}
    \label{tab:ClassesRatingvsScoreSPD}
    \begin{tabular}{lcl}
    \toprule
        \textbf{Classe} & \textbf{Code Smell} & \textbf{Bug} \\
    \midrule
        SurfaceScene & 10 & 8 \\
        Combo & 25 & 1 \\
        ElementalStrike & 15 & 1\\
        ShadowClone & 10 & 0\\
    \bottomrule
    \end{tabular}
\end{table}

The four classes presented various violations of SonarQube rules, some of the type Maintainability's Code Smell and Bugs related to Reliability, as listed in Table \ref{tab:ClassesRatingvsScoreSPD}. The estimated time for correction of the violations was 27h30m, 6h, 4h36m, and 26h for the respective four classes. Some of the rules point to aspects cited by the LLM, such as long methods\footnote{https://rules.sonarsource.com/java/RSPEC-6541/} and high cyclomatic complexity\footnote{https://rules.sonarsource.com/java/RSPEC-3776/}. The large number of estimated hours to make corrections does not impact the rating of these classes due to their size.

The two classes with an E rating and scores of 85 and 90 have few lines of code, as per Listing \ref{lst:ClasseRedButton} and \ref{lst:ClasseParalyticDart}. The \textit{RedButton} class has only one violation with a repair effort of 4h30m. The \textit{ParalyticDart} class has two violations, totaling 6h of effort. A common violation to the two classes refers to the rule \textit{Inheritance tree of classes should not be too deep}\footnote{https://rules.sonarsource.com/java/RSPEC-110/}, which points to the existence of a very deep inheritance hierarchy. This rule has a great weight in the calculation for repair cost, made by Sonar, and is invisible to the LLM, given that no information was passed beyond the source of the analyzed class. In Listing \ref{lst:ClassesRatingEScoreAltoSPD}, we have the score and respective justifications of the LLM, pointing to little or no documentation in both classes and the magic number (line 11 of Listing \ref{lst:ClasseParalyticDart}) of \textit{ParalyticDart}.

Again, it is concluded that despite overlap in some criteria, the \ac{LLM} evaluates aspects that are not evaluated by SonarQube, and its score refers to the degree of readability while SonarQube refers to the cost of repair weighted by the size of the code.

\begin{lstlisting}[style=mystyle, language=java, firstnumber=1, basicstyle=\tiny, caption={Classe RedButton (score 90)}, label={lst:ClasseRedButton}]
public class RedButton extends StyledButton {
			
	public RedButton( String label ) {
		this(label, 9);
	}

	public RedButton( String label, int size ){
		super( Chrome.Type.RED_BUTTON, label, size);
	}
	
}
\end{lstlisting}
\begin{lstlisting}[style=mystyle, language=java, firstnumber=1, basicstyle=\tiny, caption={Classe ParalyticDart (score 85)}, label={lst:ClasseParalyticDart}]
public class ParalyticDart extends TippedDart {
	{
		image = ItemSpriteSheet.PARALYTIC_DART;
	}
	
	@Override
	public int proc( Char attacker, Char defender, int damage ) {
		//when processing charged shot, only stun enemies
		if (!processingChargedShot || attacker.alignment != defender.alignment) {
			Buff.prolong(defender, Paralysis.class, 5f);
		}
		return super.proc( attacker, defender, damage );
	}
	
}
\end{lstlisting}

\begin{lstlisting}[style=mystyle, language=json, firstnumber=1, basicstyle=\scriptsize, caption={Samples of LLM's code reviews}, label={lst:ClassesRatingEScoreAltoSPD}]
RedButton: {"score":"90","reasoning":"The code is well-structured and follows good naming conventions. It also uses proper indentation and spacing. However, there is no documentation or comments to explain the purpose of the class and its methods."}
ParalyticDart: {"score":"85","reasoning":"The code follows good practices, such as proper indentation and comments. However, it lacks documentation and could benefit from more descriptive variable names. The code also has some hard-coded values that could be extracted to constants for better maintainability."}
\end{lstlisting}

We used Spearman's correlation again to compare the correlation of the other attributes with the score value of the \ac{LLM} (Table \ref{tab:tabelaCorrelacaoSpearman2}). The results obtained were similar to those of Project 1, with a significant deterioration in the identification of correlation for the \% Comments attribute, which even began to show an inverse correlation, suggesting that comments have a slight influence on worsening the score.
\begin{table}[ht] 
\footnotesize
\centering
\caption{Spearman correlation for SonarQube metrics}
\label{tab:tabelaCorrelacaoSpearman2}
\begin{tabular}{lrr}
    \toprule
    \textbf{Atributo} & \textbf{Correlação} & \textbf{Valor-p} \\
    \midrule
    Code Smells & -0.484 & 1.110e-59\\
    \% Comments & -0.149 & 2.163e-06\\
    Cog. Complexity & -0.509 & 6.917e-67\\
    Complexity & -0.536 & 2.783e-75\\
    Lines & -0.567 & 6.663e-86\\
    Statements & -0.557 & 1.836e-82\\
    \bottomrule
\end{tabular}
\end{table}
 
\subsection{\itemC}
In the study by \textcite{laskar_systematic_2023}, it was observed that newer versions of the ChatGPT model do not necessarily lead to better results. At the time of this research, the latest version of the model, called ChatGPT 4o, was released. Analyses were carried out with both projects using this version.

\subsubsection{\itemCUm}
The data compared from GPT 3.5 and 4o (Table \ref{tab:tabelaScore35vs4oQuarkus}) show a greater tendency of GPT 4o to assign higher scores to the evaluated classes. As 95.49\% (Table \ref{tab:tabelaManutenibilidadeAmostraQuarkus}) of the classes have an A rating assigned by Sonar, GPT 3.5 presented 97.36\% of the classes with a score of 80 or higher, while GPT 4o presented 99.22\%.

\begin{table}[ht]
\begin{minipage}{.4\linewidth}
    \centering
    \scriptsize
    \caption{Scores GPT3.5 and GPT4o compared (Quarkus)}
    \label{tab:tabelaScore35vs4oQuarkus}
    \begin{tabular}{ccc}
        \toprule
        & \multicolumn{2}{c}{\textbf{Classes}} \\
        \midrule
        \textbf{Score} & \textbf{GPT3.5} & \textbf{GPT4o} \\
        95 & 0 & 107 \\
        90 & 176 & 227 \\
        85 & 421 & 305 \\
        80 & 30 & 0 \\
        75 & 10 & 1 \\
        70 & 4 & 1 \\
        60 & 1 & 1 \\
        50 & 2 & 0 \\
        10 & 0 & 2 \\
        \bottomrule
    \end{tabular}
\end{minipage}
\begin{minipage}{.57\linewidth}
    \centering
    \scriptsize
    \caption{Sonar vs GPT 4o (Quarkus)}
    \label{tab:distribuicaoQuarkusSonarScore4o}
    \begin{tabular}{crrrr}
        \toprule
        \textbf{LLM} & \multicolumn{4}{c}{\textbf{Sonar Rating}} \\
        \cmidrule{2-5}
        \textbf{Score} &  A &  B &  C &  D \\
        10 &    0 &    2 &    0 &    0 \\
        60 &    1 &    0 &    0 &    0 \\
        70 &    1 &    0 &    0 &    0 \\
        75 &    1 &    0 &    0 &    0 \\
        85 &  285 &   15 &    4 &    1 \\
        90 &  220 &    4 &    2 &    1 \\
        95 &  107 &    0 &    0 &    0 \\
        \bottomrule
    \end{tabular}
\end{minipage}
\end{table}

In Table \ref{tab:distribuicaoQuarkusSonarScore4o}, the two classes that GPT 3.5 assigned a score of 50 were evaluated with a very low score by GPT 4o (score 10). The same classes considered of low quality by Sonar (rating D), which received a score of 85 by GPT 3.5, now received scores of 85 and 90 respectively. The cases of divergence were maintained, but a change in the score assigned by the LLM is noted, penalizing empty classes more severely.

\subsubsection{\itemCDois}
\captionsetup[table]{skip=1pt}
\begin{table}[ht]
\begin{minipage}{.4\linewidth}
\centering
\scriptsize
\caption{Scores GPT3.5 and GPT4o compared (SPDungeon)}
\label{tab:tabelaScore35vs4oPixel}
\begin{tabular}{ccc}
    \toprule
    & \multicolumn{2}{c}{\textbf{Classes}} \\
    \midrule
    \textbf{Score} & \textbf{GPT3.5} & \textbf{GPT4o} \\
    90 & 1& 1\\
    85 & 107& 990\\
    80 & 301&0\\
    75 & 467&5\\
    70 & 117&0\\
    65 & 0&1\\
    60 & 3&0\\
    50 & 1&0\\
    \bottomrule
\end{tabular}
\end{minipage}
\begin{minipage}{.57\linewidth}
    \centering
    \scriptsize
    \caption{Sonar vs GPT 4o (SPDungeon)}
    \label{tab:distribuicaoPBDungeonSonarScore4o}
    \begin{tabular}{crrrrr}
        \toprule
        \textbf{LLM} & \multicolumn{5}{c}{\textbf{Sonar Rating}} \\
        \cmidrule{2-6}
        \textbf{Score} &  A &  B &  C &  D & E \\
        65 &    0 &    1 &    0 &    0 & 0 \\
        75 &    4 &    0 &    1 &    0 & 0 \\
        85 &  665 &  108 &   85 &  103 & 29 \\
        90 &    1 &    0 &    0 &    0 & 0 \\
        \bottomrule
    \end{tabular}
\end{minipage}
\end{table}

The data compared from GPT 3.5 and 4o (Table \ref{tab:tabelaScore35vs4oPixel}) again demonstrate a greater tendency of GPT 4o to assign higher scores to the evaluated classes, with 99.29\% of the classes receiving a score of 85.

However, only 67.20\% (Table \ref{tab:tabelaManutenibilidadeAmostraShattered}) of the classes have an A rating assigned by Sonar, GPT 3.5 presented 41.02\% of the classes with a score of 80 or higher, while GPT 4o presented a concentration of 99.39\%. The results obtained with GPT 4o suggest that the readability and overall quality of the Shattered Pixel Dungeon project code is comparable or even superior to that of the Quarkus project. These results diverge from the evaluation made by SonarQube and the evaluation made by GPT 3.5.

When detailing the distribution of the \ac{LLM} scores for each Sonar Rating, in Table \ref{tab:distribuicaoPBDungeonSonarScore4o}, we note the existence of a low score assigned by the \ac{LLM} for a class with a high rating. There is one occurrence for rating B with a score of 65. The same \textit{AttackIndicator} class received a score of 70 from GPT 3.5, has 208 lines, and 5 Code Smells pointed out by SonarQube, totaling 4h47min of correction time. Again, some of the Code Smells were identified by the \ac{LLM} and SonarQube, such as the FIXME comment, but others were identified only by SonarQube or the \ac{LLM}. Regarding the classes with a low rating (D and E), the \ac{LLM} assigned 100\% of the scores at 85, GPT 3.5 had pointed out 3 classes with a score of 75 and 2 with a score of 70 (Table \ref{tab:distribuicaoSPDungeonSonarScore35}). The \ac{LLM} was not able to differentiate the classes considered of lower quality according to Sonar and GPT 3.5 Turbo.

We conclude that GPT 4o presented results inferior to version 3.5 Turbo when executing the prompt evaluating the readability and overall quality of the code. This degradation confirms the findings of \textcite{laskar_systematic_2023}.

\section{Limitations and future work} \label{sec:limitations}

This study conducted an evaluation on two open-source software (OSS) projects in the Java language. The number of evaluated projects may not be representative of the software industry. To mitigate this risk, large-scale projects with a large volume of code for analysis were selected. One project with a SonarQube maintainability rating of A and another B were selected. No large-volume code projects with a C or lower rating were found. Future work could conduct an extensive study with a larger volume of OSS projects.

The Java language is widely adopted in the industry, making the findings of this research of broad interest. However, similar work needs to be carried out for other languages in the future.

The comparison base was Sonar, a tool widely adopted by the industry, however, there are other static analysis tools that provide maintainability-related metrics. Different results may be obtained when using other tools as a comparison base. This would occur if the criteria used to define maintainability diverge greatly between the tools. Future work could compare the results with \ac{LLM} and other static analysis tools. The results can also be compared with other indices such as SQALE \cite{letouzey_sqale_2016} used by Sonar, the TIOBE Quality Indicator (TQI) \cite{jansen_tiobe_2023}, SIG Maintainability\cite{heitlager_practical_2007}, CISQ \cite{noauthor_list_2019}, and ISO/IEC 5055:202 \cite{iso5055_isoiec_2021}.

A change in results was identified with different versions of the \ac{LLM} model used (GPT 3.5 and 4o), confirming the finding made by \textcite{laskar_systematic_2023}. This factor limits the use of readability evaluations with \ac{LLM} for benchmarks, being valid only to compare evaluations made by the same model and the same prompt. It is necessary to carry out an extensive comparative study between various LLMs and their results in code quality analysis, since other research \cite{mohajer_skipanalyzer_2023} obtained better results in more recent versions of ChatGPT.

This study used the zero-shot technique. Different results may be obtained with the few-shot approach. Both techniques allow a model to perform tasks for which it was not explicitly trained \cite{wang_large_2023}. Future research could investigate LLMs in conversational mode, to verify the accuracy and ability to analyze code that exceeds the context window as well as its evaluation of code from related classes, as proposed by \textcite{xia_conversational_2023} in their approach for APR.

This study compared the results obtained with the ChatGPT 3.5 turbo model. ChatGPT models are the most used in research, as identified by \textcite{hou_large_2024}, with ChatGPT covered in 72 articles, ChatGPT 3.5 in 54, and ChatGPT 4 in 53. However, the literature review indicates an ecosystem of 45 variations of models and versions used in applied software engineering research since 2017. We understand the need to carry out research that compares various models and \ac{LLM} architectures, in terms of their ability to assess readability and overall code quality.

LLMs present variability of responses for the same prompt. For the context of the current work, previous studies addressed and analyzed this variability, its impact, and strategies to deal with it\cite{hao_ev_2023}. We consider it important to evaluate the performance of various prompts in evaluating the application scenario proposed in this study.

For static analyses directed at specific problems, there are studies on prompt optimization\cite{hao_ev_2023}. This work evaluated the quality of code more broadly, yet limited to characteristics detectable by static analysis tools. It used researched practices of prompt engineering \cite{white_prompt_2023}, but it is necessary to carry out studies for cost optimization in the use of \ac{LLM}. \ac{LLM} cost can be impeditive for its usage as a static analysis tool.

The results showed that SonarQube rate code quality based on the correcting cost of bad practices while hypothetically the \ac{LLM} seems to rate the code based on the easy of reading. That hypotheses deserves a dedicated research. It is also needed a research dedicated to evaluate the execution time for a \ac{LLM} to perform source code analysis and its feasibility of usage in industry constraints.

The \ac{LLM} was capable of emit a score for the source code, limited to its context window. This impose a limit to the size of source code that can be analysed by a \ac{LLM}. Future research is needed to investigate alternatives approaches to circumvent this limitation

\section{Conclusion} \label{sec:conclusion}

This work explored the use of Large Language Models (LLMs) as a tool to evaluate code quality, comparing it to a static analysis tool. Two open-source software (OSS) projects were comparatively analyzed using SonarQube and two versions of the LLM GPT, covering a total of 1,641 classes.


We identified that ChatGPT 3.5 Turbo showed a correlation of its results with the maintainability score of SonarQube, demonstrating that this version of the \ac{LLM} has the ability to evaluate the quality of a code, similarly to SonarQube. By analyzing the discrepancies, we found that the metric generated by the \ac{LLM} corresponds to the readability and current quality of the code, being limited to the evaluation of the source code provided in the prompt. In contrast, the SonarQube maintainability rating refers to the estimated repair cost of the identified bad practices. SonarQube evaluates a broader set of rules, including relationships between classes (e.g., inheritance hierarchy depth). This cost is further weighed against the estimated cost of recreating the code from scratch. This is better detailed in section \ref{sec:methodology}. Thus, classes with few lines of code have low tolerance to violations, receiving a low score from SonarQube if there is a violation that costs a few hours to correct, exceeding the creation cost of the analyzed class. The same class can receive a high score from the \ac{LLM}, if the found violation is not detectable by the \ac{LLM} or does not impact severely the readability or quality of the code.

The version of ChatGPT 4o presented inconsistent results when compared with its other evaluated version and with SonarQube result's. Leading to the conclusion that satisfactory results in one version of the \ac{LLM} do not guarantee their occurrence in future versions of the model.

Limitations such as the cost per token and the variability of responses for the same prompt are still a challenge for the use of LLMs in substitution to static analysis tools to evaluate code quality. However, there is potential for their use in complement to these tools in order to observe quality characteristics not captured by them and enabling the creation of new quality indices.

This exploratory study evidences that LLMs can be used to evaluate the quality of source code, and can even expand the analysis currently carried out by static analysis tools by covering aspects that are currently not captured by them. New studies are necessary to investigate these different aspects of quality, as well as to propose mechanisms to deal with the cost and variability of outputs produced by LLMs.

\printbibliography

\end{document}